\documentclass[prl,twocolumn]{revtex4}
\usepackage{graphicx}
\begin{document}
\title{Atom-molecule coherence in a one-dimensional system}
\author{R. Citro}
\affiliation{ Dipartimento di Fisica ``E. R. Caianiello'' and
Unit{\`a} I.N.F.M. di Salerno\\ Universit{\`a} degli Studi di Salerno, Via S.
Allende, I-84081 Baronissi (Sa), Italy}
\author{E. Orignac}
\affiliation{Laboratoire de Physique Th\'eorique de l'\'Ecole Normale
  Sup\'erieure CNRS-UMR8549 \\ 24, Rue Lhomond  F-75231 Paris Cedex 05
  France}
\begin{abstract}
We study a model of one-dimensional fermionic atoms with a narrow
Feshbach resonance  that allows them to bind in
pairs to form bosonic molecules.  We show
that at low energy, a coherence develops between the molecule and
fermion Luttinger liquids. At the same time, a gap opens in the spin
excitation spectrum. The coherence implies that the order parameters
for the molecular Bose-Einstein Condensation
and the atomic BCS pairing become identical. Moreover,
 both bosonic and fermionic charge density wave correlations decay
exponentially, in contrast with a usual Luttinger liquid.
 We exhibit a  Luther-Emery point where the systems can be described
 in terms of  noninteracting pseudofermions. At this point, we provide
 closed form expressions for the density-density response functions.
\end{abstract}
\maketitle

Recent experiments on fermionic ${}^6$Li or ${}^{40}$K atoms in
optical traps have led to the realization of paired superfluidity.
Pairs of atoms were found to bind into bosonic molecules that
displayed a Bose-Einstein
condensation\cite{jochim03_mloecules_bec,greiner_bec,zwierlein_bec,strecker_bec}
as the magnetic field was varied across a Feshbach resonance. The
Bose condensation of the molecules is expected to trigger
superfluidity of the fermions. A crossover is thus expected from a
BCS superfluid of paired atoms to a Bose Einstein condensate (BEC)
of molecules  as the molecules become more tightly bound. It is
important to stress that since the  BEC and the BCS states have
the same broken symmetry, there is no fundamental distinction
between them, and these two extreme limits are connected by a
smooth crossover. This crossover is naturally described by the
fermion-boson model which has extensively been considered in the
context of bipolaronic and high-Tc
superconductivity\cite{bosefermi_htc} and has
recently known a regain of theoretical interest
\cite{timmermans01_bosefermi_model,holland01_bosefermi_model,ohashi03_transition_feshbach,ohashi03_collective_feshbach}.

Recently, the BCS-BEC crossover has been investigated in the
one-dimensional
case\cite{fuchs04_resonance_bf,fuchs_bcs_bec,tokatly_bec_bcs_crossover1d}.
Despite the absence of broken symmetries in one dimension, a rich
phenomenology is known to emerge such as collectivization of
single-particle degrees of freedom, spin-charge separation and
quasi-long range order\cite{giamarchi_book_1d}. In particular,
spin-1/2 fermions with attractive interactions give rise to a
state with gapless charge degrees of freedom and gapped spin
degrees of freedom, with quasi-long range superconducting and
charge density wave order, known as the Luther-Emery
liquid\cite{giamarchi_book_1d}. In
\cite{fuchs_bcs_bec,tokatly_bec_bcs_crossover1d}, an integrable
 model\cite{gaudin_yang} of spin-1/2
fermions with attractive interactions was considered. It was shown
that the Luther-Emery liquid obtained at small attraction was
crossing over to a Luttinger liquid of tightly bound bosonic
molecules. In \cite{fuchs04_resonance_bf}, the boson-fermion model
was used, but the case of a broad resonance
 where only the bosons or only the fermions are present was
considered, yielding results  similar to those of the
integrable model. An interesting theoretical question is to
understand what happens in the case of a narrow Feshbach resonance
when the atoms and the molecules can coexist.
 In the present Letter, we investigate such a case. We show
that a strongly correlated state exists in this case, in which the
order parameters of the Bose condensation and superfluidity decay
with the same critical exponent, and density fluctuations near the
Fermi wavevector are strongly suppressed. Also, a spectral gap is
present in the atomic spectral function, as in the case of a
Luther-Emery liquid. From the experimental point of view, the
interest in working in one dimension is the possibility of using
the confinement induced resonance (CIR)
\cite{olshanii_yurovsky} to form the molecules, as
recently demonstrated\cite{moritz05_molecules1d}. Another way of
forming the molecules is by photoassociation
techniques\cite{mackie_photoassociation}. Since it is possible to
realize quasi-1d systems of bosons and fermions in optical
traps\cite{petrov04_bec_review}, some of our predictions could be
testable in experiments.

We consider a 1D system of  fermionic atoms that can bind reversibly
into bosonic molecules. In  the continuum case, the
Hamiltonian of the system reads:
\begin{eqnarray}
  \label{eq:nolattice}
  H=&&-\int dx \sum_\sigma  \psi^\dagger_\sigma \frac {\nabla^2}{2m_f}
  \psi_\sigma   - \int dx \psi_b^\dagger  \left(\frac  {\nabla^2}
  {4m_f} -\nu\right)
  \psi_b \nonumber \\
&&+  \lambda \int dx (\psi_b^\dagger  \psi_\uparrow
  \psi_\downarrow + \psi^\dagger_\downarrow  \psi^\dagger_\uparrow
  \psi_b) \nonumber \\ && + \frac 1 2 \sum_{\nu,\nu'=b,f} \int dx dx' V_{\nu \nu'}(x-x') \rho_\nu(x) \rho_{\nu'}(x')
\end{eqnarray}
where $\psi_\sigma$ annihilates a fermion (atom) of spin $\sigma$,
$\psi_b$ annihilates a boson (molecule),
$\rho_b=\psi^\dagger_b\psi_b$
$\rho_f=\sum_\sigma \psi^\dagger_\sigma\psi_\sigma$, and we have set
$\hbar=1$.  The parameters
  $V_{ff}$, $V_{bb}$ and $V_{bf}$ measure (respectively) the fermion-fermion,
  boson-boson, and fermion-boson repulsion. The parameter $\nu$ is the
  detuning, and the term
  $\lambda$ describes the binding of  pair of  atoms into a
   molecule and the reverse process. When $\lambda\ne 0$,
  only the total atom number ${\cal N}=2N_b+N_f$ is conserved.
 The fermion-boson model (\ref{eq:nolattice})
 can be analyzed  by bosonization techniques in the limit
  of a narrow resonance (i.e. small $\lambda$)  and provided that neither the
  density of atoms nor the density of molecules vanishes.

 First, we recall the bosonized description of the
  system when boson-fermion interactions are turned off.
For $\lambda=0,V_{bf}=0$, both the number of free atoms $N_f$ and
the
  number of molecules $N_b$ are conserved and the Hamiltonian
  equivalent
to (\ref{eq:nolattice}) is given by\cite{giamarchi_book_1d}:
  \begin{eqnarray}
    \label{eq:bosonized-decoupled}
    H&=&H_b+H_\rho+H_\sigma -\frac{2g_{1\perp}}{(2\pi\alpha)^2}
    \int dx
    \cos \sqrt{8}\phi_\sigma \\
     \label{eq:bosonized-bosons}
    H_\nu&=&\int \frac{dx}{2\pi} \left[ u_\nu K_\nu (\pi \Pi_\nu)^2 +\frac {u_\nu}
    {K_\nu}(\partial_x\phi_\nu)^2\right]
\end{eqnarray}
\noindent where
    $[\phi_{\nu}(x),\Pi_{\nu'}(x')]=i\delta(x-x')\delta_{\nu,\nu'}$,
    ($\nu,\nu'=b,\sigma,\rho$).
The parameters $u_\rho,u_\sigma,u_b$ are the velocity of respectively
    atomic density, atomic spin,
    and molecular density excitations.
    $K_\rho$ and $K_b$ are the Luttinger
    exponents\cite{giamarchi_book_1d}. They decrease as
(respectively) the atom-atom and
molecule-molecule repulsion increase. For weak repulsion, it is
    possible to express $K_\rho$ as a function of  the scattering
    length $a_s$, trapping frequency  $\omega_\perp$ and the Fermi
    velocity of non-interacting atoms as
$K_\rho=(1+2 a_s\omega_\perp/\pi v_F)^{-1/2}$. In
the case of bosons, the Luttinger exponent $K_b$ must be extracted
from the Lieb-Liniger
equations\cite{lieb_bosons_1D,giamarchi_book_1d}.
For weak interaction, $K_b \to +\infty$, while in the Tonks-Girardeau
    limit, $K_b=1$. Turning to the spin interaction, it is
    known\cite{giamarchi_book_1d} that
under the renomalization group (RG)
 $g_{1\perp}$ and $K_\sigma$ flow to the fixed point
    $K_{\sigma}^*=1,g_{1\perp}^*=0$ provided that the repulsive
    interactions respect the SU(2) spin symmetry. We will thus replace
    $K_\sigma, g_{1\perp}$ by their fixed point value in the
    following. The power of bosonization in treating interacting
    systems in one-dimension comes from the possibility of expressing
    the fermion and boson annihilation operators in terms of the
    fields in (\ref{eq:bosonized-decoupled}) as:
    \begin{eqnarray}
      \label{eq:fermion-bosonized}
&&      \psi_{r,\sigma}(x)
=\frac{e^{\frac{i}{\sqrt{2}}
      [\theta_{\rho}-r\phi_{\rho} +\sigma
      (\theta_{\sigma}-r\phi_{\sigma})](x)}}{\sqrt{2\pi\alpha}}, \\
  \label{eq:boson-bosonized}
&&   \Psi_b(x)=\frac {e^{i\theta_b}}{\sqrt{2\pi\alpha}} \left[ 1 +
A \cos (2\phi_b -2 k_B x) \right],
\end{eqnarray}
\noindent where
  the dual fields\cite{giamarchi_book_1d}
 $\theta_\nu(x) =\pi \int^x \Pi_\nu(x')dx'$
($\nu=b,\rho,\sigma$) have been introduced, $k_F=\pi N_f/L$,
  $k_B=\pi N_b/L$, $r=\pm$, and $\alpha$ is a cutoff.
The atom field is expressed in terms of the right and left moving
 fields of Eq.~(\ref{eq:fermion-bosonized}) as:
 $\psi_\sigma(x)=e^{ik_F x} \psi_{+,\sigma}+ e^{-ik_F x}
 \psi_{-,\sigma}$.
Similarly, the atom and molecule density are
given by\cite{giamarchi_book_1d}:
\begin{eqnarray}
  \label{eq:fermion-density-bosonized}
  \rho_f(x)&=&-\frac{\sqrt{2}}{\pi}\partial_x\phi_\rho +\frac{\cos (2k_F
  x -\sqrt{2}\phi_{\rho})}{\pi \alpha}\cos \sqrt{2}\phi_\sigma,\\
  \label{eq:boson-density-bosonized}
  \rho_b(x)&=&-\frac{1}{\pi}\partial_x\phi_b +\frac{\cos (2k_B
  x - 2\phi_b)}{\pi \alpha}.
\end{eqnarray}
Let us now turn on  $\lambda,V_{bf},\nu \ne 0$, assuming that
that they are small compared with the kinetic energy of the atoms and
the molecules. This
corresponds to the
    limit of a narrow Feshbach resonance\cite{fuchs04_resonance_bf}.
 For $k_B\ne
  k_F$, the boson-fermion repulsion reduces to a term $
\sim \frac{V\sqrt{2}} {\pi^2} \int \partial_x \phi_b \partial_x
\phi_{\rho}$\cite{cazalilla_mathey}.
Using Eqs.~(\ref{eq:fermion-bosonized})-(\ref{eq:boson-bosonized}) The
bosonized form of the $\lambda$ is:
\begin{eqnarray}
  \label{eq:lambda-bosonized}
H_{bf}=\frac{\lambda}{\sqrt{2\pi^3} \alpha } \int dx \cos (\theta_b
  -\sqrt{2}\theta_\rho) \cos \sqrt{2}\phi_{\sigma}
\end{eqnarray}
Finally, the detuning term reads: $-\frac{\nu}{\pi}\int dx
\partial_x\phi_b$,  which shows that is
can be eliminated by a shift of $\phi_b \to \phi_b -2\frac{\nu x
  K_b}{u_b}$, without affecting Eq.~(\ref{eq:lambda-bosonized}).
Thus, the detuning is effective only when it induces
band-filling transitions at which either the density of
  atoms ($\nu<0$) or of molecules ($\nu>0$) vanishes.
At these transitions, the bosonization description breaks down,
with  divergence of the compressibility.\cite{cabra_instabilityLL}
The large
detuning  limit has been analyzed in \cite{fuchs04_resonance_bf}, where it
was shown that existence of virtual atom or molecule states only leads to a
renormalization of the Luttinger exponent $K_b$ or $K_\rho$ respectively.
In the following, we take $\nu=0$.
The RG equation for $\lambda$ reads:
  \begin{eqnarray}
    \frac{d\lambda}{dl} = \left(\frac 3 2 -\frac 1 {2K_{\rho}} -\frac
    1 {4 K_b}\right) \lambda,
  \end{eqnarray}
 showing that for $\frac 1 {2K_{\rho}} -\frac 1 {4 K_b} <3/2$,
this interaction is relevant. Since for hard core
bosons\cite{schultz_1dbose} $K_b=1$ and for free bosons
$K_b=\infty$  while for free fermions $K_\rho=1$,
this inequality is satisfied except for very strongly
repulsive interactions $V_{ff}$ and $V_{bb}$.
In this strong repulsion limit, there is no coherence between atoms
and molecules and the
system is described by the theory of
\cite{cazalilla_mathey}. For less repulsive
interactions, the relevance of $\lambda$ drives the system to a new
fixed point. To understand the nature of this fixed point,  it is
convenient  to perform a canonical
  transformation\cite{giamarchi_book_1d}
 $\theta_{-}=\frac 1 {\sqrt{3}}\theta_b-
\frac{\sqrt{2}}{\sqrt{3}}\theta_\rho$,
$\theta_{+}=\frac{\sqrt{2}}{\sqrt{3}}\theta_b+\frac 1
{\sqrt{3}}\theta_\rho $ and the same transformation for the
$\phi_\nu$. Then, $H_b+H_\rho$ becomes
$H_\pm$ with:
\begin{eqnarray}
\label{eq:b-plus-rho}
  H_{\pm}&=&\int \frac{dx}{2\pi}\sum_{\nu=\pm} \left[ u_\nu K_\nu
  (\pi \Pi_\nu)^2 + \frac{u_\nu}{K_\nu}(\partial_x \phi_\nu)^2\right]
  \nonumber \\ && + \int \frac{dx}{2\pi} [g_1 (\pi \Pi_+)(\pi \Pi_-) + g_2
  \partial_x\phi_+ \partial_x\phi_-],
\end{eqnarray}
\noindent where $u_\pm,K_\pm,g_{1,2}$ can be obtained in terms of the
  parameters of the original Hamiltonian and $V_{bf}$. In the following, we
  will assume $u_\pm=u_\sigma=u$.
When we express (\ref{eq:lambda-bosonized}) in terms of
$\theta_-$,
we see that when $\lambda$ is relevant, only the
 fields  $\phi_\sigma$ and
 $\theta_-$ are locked, with a gap $\Delta_-=\Delta_\sigma\sim
 u/\alpha (\lambda \alpha/u)^{4/(6-2K_\rho^{-1}-K_b^{-1})}$, while $\phi_+$
 remains gapless.
 The field $\phi_+$ describes the total density excitations of the
 system, as can be seen from the relation
$\sqrt{6}[\phi_+(\infty)-\phi_+(-\infty)]={\cal N}$. It is described
 by the low-energy Hamiltonian:
\begin{eqnarray}\label{eq:gapless-modes}
  H_+=\int \frac{dx}{2\pi} \left[ u^*_+ K^*_+ (\pi \Pi_+)^2 +\frac {u^*_+}
    {K^*_+}(\partial_x\phi_+)^2\right],
\end{eqnarray}
where $u^*_+,K^*_+$ are renormalized values of $u_+,K_+$, resulting
from the  residual
interactions with the gapped mode $\theta_-$ caused by the terms $g_1,g_2$.
 The gapful excitations are formed by kinks of the fields
 $\theta_-$ and $\phi_\sigma$ such that
 $\theta_-(\infty)-\theta_-(-\infty)=\pm \pi/\sqrt{3}$ and
 $\phi_{\sigma_+}(\infty) -\phi_{\sigma_-}(-\infty)=\pm
 \pi/\sqrt{2}$.
As a result, they carry a spin $\pm 1/2$ and create  a phase difference of  $\mp
 \pi$ between bosons and fermions.  Thus, they can be identified with
 half-vortices carrying a spin $1/2$.  Having understood the spectrum of
 the system, let us turn to its ground state correlations.
The locking of the fields $\theta_-$ and
$\phi_\sigma$ yields from Eq.~(\ref{eq:boson-bosonized}) the following
low energy expression for the BEC order parameter for the molecules:
    $\Psi_B(x)\sim e^{i\sqrt{\frac 2 3} \theta_+}$,
while the order parameter for s-wave BCS superfluidity of the fermions
becomes: $\psi_\uparrow \psi_\downarrow \sim e^{i\sqrt{2}\theta_\rho} \cos
  \sqrt{2}\phi_\sigma \sim e^{i\sqrt{\frac 2 3} \theta_+}$.
Thus,  these two  order parameters become
\emph{identical} in the low energy
limit as in the case of higher
dimensionality.\cite{ohashi03_transition_feshbach,ohashi03_collective_feshbach}.
The boson correlator behaves as:
 $\langle \Psi_B^\dagger(x,\tau) \Psi_B(0,0)\rangle =((x^2 +u^2
  \tau^2)/\alpha^2)^{-\frac 1 {6K_+}}$, yielding a molecule momentum
distribution  $n_B(k)\sim |k|^{1/(3K_+)-1}$.  This momentum
distribution could be measured in a condensate expansion
experiment\cite{gerbier05_phase_mott_coldatoms,altman04_exploding_condensates}.
A more striking consequence of the locking $\theta_-$ is
that both the $\pm 2k_B$ and the $\pm 2k_F$ harmonic in  (respectively)
$\rho_b(x)$ and $ \rho_F(x)$  acquire exponentially  decaying
correlations, with a correlation
length $\sim u/\Delta_-$. The origin of
this exponential decay is that both $\rho_b$ and $\rho_f$ depend on
the field $\phi_-$ dual to $\theta_-$\cite{giamarchi_book_1d}.
Such behavior is
in contrast with the behavior of a Luttinger liquid of molecules, or a
Luther Emery liquid of fermions with attractive interaction, in which
these correlation functions would decay as power law.

A more detailed picture of the gapful spectrum and the correlation
functions can be obtained at a particular solvable point of the
parameter space, the so called Luther-Emery
point\cite{luther_exact}. For $K_-=3/2$,
 one can introduce the pseudofermion fields:
\begin{eqnarray}
 \Psi_{r,\sigma}=\frac {e^{i[(\sqrt{\frac 2 3}\phi_- -r \sqrt{\frac 3
  2}\theta_-)  +
  \sigma(\theta_\sigma-r \phi_\sigma)]}}{\sqrt{2\pi\alpha}},
\end{eqnarray}
\noindent to rewrite Eq.~(\ref{eq:lambda-bosonized}) and
  $H_-+H_\sigma$ as a free
  pseudofermions Hamiltonian:
\begin{eqnarray}\label{eq:fermionized-ham}
  H= \sum_\sigma \int dx \left[ -iu \sum_{r=\pm}
   r \Psi_{r,\sigma}^\dagger \partial_x \Psi_{r,\sigma} +
   \frac{\lambda}{\sqrt{8\pi}}  \Psi_{r,\sigma}^\dagger
   \Psi_{r,\sigma}\right].
\end{eqnarray}
  At that point, the kinks are pseudofermions with dispersion
   $\epsilon(k)=\pm \sqrt{(u k)^2 +\Delta^2}$, where
   $\Delta=\frac{|\lambda|}{\sqrt{8\pi}}$.
In terms of the pseudofermions the fermion density reads:
  $\rho_{2k_F,f}(x) =e^{i\left[\sqrt{\frac 2 3} \phi_+ -2k_F x\right]} (  \Psi^\dagger_{-,\uparrow}
  \Psi^\dagger_{+,\downarrow}  +\Psi^\dagger_{+,\uparrow}
   \Psi^\dagger_{-,\downarrow} ) + \text{H. c.}$  which yields an
   expression of density-density correlations correlation in terms of
   modified Bessel functions.  The boson density is
expressed in terms of order and disorder operators of four  2D non
critical Ising
models\cite{luther_ising} as: $\rho_b(x)\sim e^{i(\sqrt{8/3} \phi_+ -
  2k_B x)} (\mu_{1,\uparrow}\sigma_{2,\uparrow} +
  i \sigma_{1,\uparrow} \mu_{2,\uparrow})  (\mu_{1,\downarrow}\sigma_{2,\downarrow} +
  i\sigma_{1,\downarrow} \mu_{2,\downarrow}) +\text{H. c.} $
where $\sigma(\mu)$ are the order (disorder) parameters of the
Ising model. The correlation functions of the non-critical Ising model
   are obtained from \cite{wu_ising_correlations}. The Fourier
   transform of the Matsubara correlation functions then reads:
\begin{widetext}
\begin{eqnarray}\label{eq:boson_structure_factor}
  \chi_{\rho\rho}^B(\pm 2k_B+q,\omega)&=&\frac {2\pi}{u}
\left(\frac{\Delta\alpha}{u}\right)^{\frac{4K_+}{3}}
\left(\frac \Delta u\right)^2
  \frac{\sqrt{\pi}\Gamma\left(1-\frac{2K_+}{3}\right)^3}{4\Gamma\left(\frac
  3 2 -\frac{2K_+}{3}\right)}
  {}_3F_2\left(1-\frac{2K_+}{3},1-\frac{2K_+}{3},1-\frac{2K_+}{3};\frac
  3 2 -\frac{2K_+}{3},1;-\frac{\omega^2+(uq)^2}{4\Delta^2}\right), \\
\label{eq:fermion_structure_factor}
  \chi_{\rho\rho}^F(\pm 2k_F + q,\omega)&=&\frac 1 {2\pi u}
  \left(\frac{\Delta\alpha}{u}\right)^{\frac{K_+}{3}} \left[
  \frac{\Gamma\left(1-\frac{K_+}{6}\right)^3}{\Gamma\left(\frac 3 2
  -\frac{K_+}{6}\right)}
  {}_3F_2\left(1-\frac{K_+}{6},1-\frac{K_+}{6},1-\frac{K_+}{6};\frac 3
  2
  -\frac{K_+}{6},1;-\frac{\omega^2+(uq)^2}{4\Delta^2}\right)\right.\nonumber
  \\ &&\left. + \frac{\Gamma\left(2-\frac{K_+}{6}\right)\Gamma\left(1-\frac{K_+}{6}\right)\Gamma\left(-\frac{K_+}{6}\right)}{\Gamma\left(\frac 3 2
  -\frac{K_+}{6}\right)}   {}_3F_2\left(2-\frac{K_+}{6},1-\frac{K_+}{6},-\frac{K_+}{6};\frac 3
  2 -\frac{K_+}{6},1;-\frac{\omega^2+(uq)^2}{4\Delta^2}\right) \right],
\end{eqnarray}
\end{widetext}
\noindent and the response functions are then obtained by the substitution
$i\omega \to \omega+i0$. Since the generalized hypergeometric
functions ${}_{p+1}F_p(\ldots;\ldots;z)$ are analytic for $|z|<1$
\cite{erdelyi_functions_1}, the imaginary part of the response
functions vanishes if $\omega<2\Delta$. For $\omega>2\Delta$, the
behavior of the imaginary part is given by the
expression\cite{olsson66_gen_hypergeometric}
of the imaginary part of ${}_3F_2$  in terms
of Appell's hypergeometric function\cite{erdelyi_functions_1}
$F_3$, yielding power law singularities at $\omega=2\Delta$
with an exponent $K_+/3-1/2$ for
the atoms and $4K_+/3-1/2$ for the molecules. The Luther-Emery limit
also yieds the $q\simeq 0$ components of the density response.
Noticing that $\rho_B = \partial_x
\phi_+/\sqrt{3} + \sqrt{2} \partial_x \phi_-/\sqrt{3}$, we find that
$\Im \chi_{\rho\rho}^B(q,\omega)$ is the sum of a term
$\propto \delta(\omega \pm u_+ q)$ coming from the gapless phase mode,
and a term $\propto (\omega -2\Delta - (uq)^2/(4\Delta^2))^{-1/2}
\Theta(\omega -2\Delta - (uq)^2/(4\Delta^2))$ coming from the gapped
mode. The same result holds for $\Im
\chi_{\rho\rho}^F(q,\omega)$. Most interestingly, the cross
correlations of the fermion and the boson density are also
non-vanishing for $q\to 0$ and behave similarly to
$\chi_{\rho\rho}^B(q,\omega)$.
The imaginary parts of correlation functions
Eq.~(\ref{eq:boson_structure_factor})--~(\ref{eq:fermion_structure_factor})
can be measured by Bragg spectroscopy with large momentum
transfer\cite{stenger99_bragg_bec}
 and the $q=0$ component can be measured by Bragg
spectroscopy with small momentum transfer\cite{stamper-kurn99_bragg_bec,zambelli00_dynamical_structure_factor,steinhauer_bec_structure_factor}.
The atom Green's function is obtained using form factor
expansion
techniques
\cite{lukyanov_soliton_ff,tsvelik_spectral_cdw}as:
\begin{eqnarray}
  G(x,\tau)\sim e^{i\varphi} \left(\frac \alpha {\rho} \right) ^{\frac 1
  {12}\left( K_++\frac 1 {K_+}\right)} K_{\frac 5 6} \left(\frac \Delta v
  \rho \right) + O(e^{-3M\rho/v}),
\end{eqnarray}
where $\rho=\sqrt{x^2 +v^2\tau^2}$.The corresponding spectral function
 is obtained by Fourier transformation\cite{tsvelik_spectral_cdw}, and it
 vanishes below the
 gap $\Delta$, as in a superfluid\cite{abrikosov_book}. It would be
 interesting to observe this gap in a condensate expansion
 experiment\cite{gerbier05_phase_mott_coldatoms,altman04_exploding_condensates}.
We have seen that with a narrow resonance, the mutual coherence of the
 atoms and the molecules
reinforces the superfluidity of the system.
An important question to ask is how such a
coherence can be lost. A first way of losing the coherence is by
applying a temperature, creating a density of half-vortices $\sim
e^{-\Delta_\sigma/T}$, which destroys phase coherence
between the atoms an molecules on a lengthscale $\ell(T)\sim
e^{\Delta_\sigma/T}$.
The second  way of losing the coherence between
 bosons and fermions is by applying a magnetic field strong enough to
 cancel the gap for the creation of half-vortices
 excitations, causing  a
commensurate incommensurate
transition.\cite{cic_transition}
As a result of this transition,  power law singularities in the
density-density correlations reappear, and
the behavior of the system
is again described by the models of
Refs.~\cite{cazalilla_mathey}.
>From the experimental point of view, a narrow resonance could be
 obtain by working with ${}^6$Li atoms\cite{strecker_bec} trapped in a
 two dimensional optical lattice. The relevant parameters\cite{strecker_bec,bruun05_6li_interaction} for
 ${}^6$Li are the mass $m$(${}^6$Li)=$9.96 \times 10^{-27}$ kg,
 the width of the resonance $\Delta B=0.23$G = $2.3\times 10^{-5}$
 T, the atom-atom scattering length $a_{bg}$=80 $a_0$ = $4.23 \times 10^{-9}$
 m, the difference in magnetic moment between the atom and the
 molecule, $\Delta\mu \sim 2\mu_B$=$2\times 927.400 949 \times
 10^{-26}=1.8\times 10^{-23}$ J.T$^{-1}$, and the trapping
 frequencies $\omega_\perp$=$2\pi\times 69$kHz, $\omega_z$=
$\omega_\perp/270$. With these parameters we find that the Fermi
velocity is of the order $10^{-2}$ m/s and the value of the
Luttinger parameter is $K_\rho\sim 0.995$. Finally we would like
to comment on the fact we have only considered the case of the
continuum system~(\ref{eq:nolattice}). All the considerations
above apply equally to a  lattice model at an incommensurate
filling, albeit the effective mass of the atoms and the molecules
can be strongly enhanced by the periodic potential. For
commensurate filling in the lattice system, an umklapp term, $
H_{umk.}=\frac{2g_U}{(2\pi\alpha)^2} \int dx \cos \sqrt{24}\phi_+$
   is present in the Hamiltonian and can create a Mott gap\cite{giamarchi_book_1d}. The RG treatment shows that the
   Mott gap exists only for $K_+<1/3$ i.e. very strong repulsion.
   In the Mott insulating state, superfluid and BEC correlations
   become short ranged as the density density correlations.

\textit{Note added:} The model considered in the present paper has
also been studied independently by D. E. Sheehy and
L. Radzihovsky\cite{sheehy_feshbach}. The CSG phase discussed by these
authors is identical to the one discussed in the present paper, and
the condition for stability of the CSG phase derived by them is also
compatible with the one derived in this paper.


\begin{thebibliography}{10}

\bibitem{jochim03_mloecules_bec}
S.~Jochim, M.~Bartenstein, A.~Altmeyer, G.~Hendl, S.~Riedl, C.~Chin, J.~Hecker Denschlag and R.~Grimm,
\newblock Science {\bf 302}, 2101 (2003).

\bibitem{greiner_bec}
M.~Greiner, C.~A. Regal, and D.~S. Jin,
\newblock Nature {\bf 426}, 537 (2003).

\bibitem{zwierlein_bec}
M.~W. Zwierlein,  C.~A. Stan, C.~H. Schunck, S.~M.~F. Raupach,
S.~Gupta, Z.~Hadzibabic, and W.~Ketterle,
\newblock Phys. Rev. Lett. {\bf 91}, 250401 (2003).

\bibitem{strecker_bec}
K.~E. Strecker, G.~B. Partridge, and R.~G. Hulet,
\newblock Phys. Rev. Lett. {\bf 91},  080406 (2003).

\bibitem{bosefermi_htc}
A.~S. Alexandrov and J.~Ranninger,
\newblock Phys. Rev. B {\bf 23}, 1796 (1981);
R.~Friedberg and T.~D. Lee,
\newblock Phys. Rev. B {\bf 40}, 6745 (1989).

\bibitem{timmermans01_bosefermi_model}
E.~Timmermans, K.~Furuya, P.~W. Milonni, and A.~K. Kerman,
\newblock Phys. Lett. A {\bf 285}, 228 (2001).

\bibitem{holland01_bosefermi_model}
M.~Holland, S.~J. J. M.~F. Kokkelmans, M.~L. Chiofalo, and R.~Walser,
\newblock Phys. Rev. Lett. {\bf 87}, 120406 (2001).

\bibitem{ohashi03_transition_feshbach}
Y.~Ohashi and A.~Griffin,
\newblock Phys. Rev. A {\bf 67}, 033603 (2003).

\bibitem{ohashi03_collective_feshbach}
Y.~Ohashi and A.~Griffin,
\newblock Phys. Rev. A {\bf 67}, 063612 (2003).

\bibitem{fuchs04_resonance_bf}
J.~Fuchs, A.~Recati, and W.~Zwerger,
\newblock Phys. Rev. A {\bf 71}, 033630 (2005).

\bibitem{fuchs_bcs_bec}
J.~Fuchs, A.~Recati, and W.~Zwerger,
\newblock Phys. Rev. Lett. {\bf 93}, 090408 (2004).

\bibitem{tokatly_bec_bcs_crossover1d}
I.~V. Tokatly,
\newblock Phys. Rev. Lett. {\bf 93}, 090405 (2004).

\bibitem{giamarchi_book_1d}
T.~Giamarchi,
\newblock {\em Quantum Physics in one Dimension}, volume 121 of {\em
  International series of monographs on physics},
\newblock Oxford University Press, Oxford, UK, 2004.

\bibitem{gaudin_yang}
M.~Gaudin,
\newblock Phys. Lett. A {\bf 24}, 55 (1967);
C.~N. Yang,
\newblock Phys. Rev. Lett. {\bf 19}, 1312 (1967).

\bibitem{olshanii_yurovsky}
M.~Olshanii,
\newblock Phys. Rev. Lett. {\bf 81}, 938 (1998);
V.~A. Yurovsky,
\newblock Phys. Rev. A {\bf 71}, 012709 (2005).

\bibitem{moritz05_molecules1d}
H.~Moritz, T.~{St\"oferle}, K.~{G\"unter}, M.~{K\"ohl}, and T.~Esslinger,
\newblock Phys. Rev. Lett. {\bf 94}, 210401 (2005).

\bibitem{mackie_photoassociation}
M.~Mackie, E.~Timmermans, R.~Cote, and J.~M. Javanainen,
\newblock Opt. Express {\bf 8}, 118 (2001).

\bibitem{petrov04_bec_review}
D.~Petrov, D.~Gangardt, and G.~Shlyapnikov,
\newblock J. Phys. IV {\bf 116}, 3 (2004).

\bibitem{lieb_bosons_1D}
E.~H. Lieb and W.~Liniger,
\newblock Phys. Rev. {\bf 130}, 1605 (1963).

\bibitem{cazalilla_mathey}
M.~A. Cazalilla and A.~F. Ho,
\newblock Phys. Rev. Lett. {\bf 91}, 150403 (2003);
L.~Mathey, D.-W. Wang, W.~Hofstetter, M.~D. Lukin, and E.~Demler,
\newblock Phys. Rev. Lett. {\bf 93}, 120404 (2004).

\bibitem{cabra_instabilityLL}
D.~C. Cabra and J.~E. Drut,
\newblock J. Phys.: Condens. Matter {\bf 15}, 1445 (2003).

\bibitem{schultz_1dbose}
T.~D. Schultz,
\newblock J. Math. Phys. {\bf 4}, 666 (1963).

\bibitem{gerbier05_phase_mott_coldatoms}
F.~Gerbier , A.~Widera, S.~{F\"olling}, O.~Mandel, T.~Gericke, and I.~Bloch,
\newblock Phys. Rev. Lett. {\bf 95}, 050404 (2005).

\bibitem{altman04_exploding_condensates}
E.~Altman, E.~Demler, and M.~D. Lukin,
\newblock Phys. Rev. A {\bf 70}, 013603 (2004).

\bibitem{luther_exact}
A.~Luther and V.~J. Emery,
\newblock Phys. Rev. Lett. {\bf 33}, 589 (1974).

\bibitem{luther_ising}
A.~Luther and I.~Peschel,
\newblock Phys. Rev. B {\bf 12}, 3906 (1975).

\bibitem{wu_ising_correlations}
T.~Wu, B.~McCoy, C.~Tracy, and E.~Barouch,
\newblock Phys. Rev. B {\bf 13}, 1976 (1976).

\bibitem{erdelyi_functions_1}
A.~{Erd\'elyi}, W.~Magnus, F.~Oberhettinger, and F.~G. Tricomi,
\newblock {\em Higher transcendental functions}, volume~1,
\newblock McGraw-Hill, NY, 1953.

\bibitem{olsson66_gen_hypergeometric}
P.~Olsson,
\newblock J. Math. Phys. {\bf 7}, 702 (1966).

\bibitem{stenger99_bragg_bec}
J.~Stenger et~al.,
\newblock Phys. Rev. Lett. {\bf 82}, 4569 (1999),
\newblock Phys. Rev. Lett. \textbf{84}, 2283(E) (2000).

\bibitem{stamper-kurn99_bragg_bec}
D.~M. Stamper-Kurn et~al.,
\newblock Phys. Rev. Lett. {\bf 83}, 2876 (1999).

\bibitem{zambelli00_dynamical_structure_factor}
F.~Zambelli, L.~Pitaevskii, D.~M. Stamper-Kurn, and S.~Stringari,
\newblock Phys. Rev. A {\bf 61}, 063608 (2000).

\bibitem{steinhauer_bec_structure_factor}
J.~Steinhauer, R.~Ozeri, N.~Katz, and N.~Davidson,
\newblock Phys. Rev. Lett. {\bf 88}, 120407 (2002).

\bibitem{lukyanov_soliton_ff}
S.~Lukyanov and A.~B. Zamolodchikov,
\newblock Nucl. Phys. B {\bf 607}, 437 (2001).


\bibitem{tsvelik_spectral_cdw}
A.~M. Tsvelik and F.~H.~L. Essler,
\newblock Phys. Rev. Lett. {\bf 90}, 126401 (2003).

\bibitem{abrikosov_book}
A.~A. Abrikosov, L.~P. Gorkov, and I.~E. Dzyaloshinski,
\newblock {\em Methods of Quantum Field Theory in Statistical Physics},
\newblock Dover, New York, 1963.

\bibitem{cic_transition}
G.~I. Japaridze and A.~A. Nersesyan,
\newblock JETP Lett. {\bf 27}, 334 (1978);
V.~L. Pokrovsky and A.~L. Talapov,
\newblock Phys. Rev. Lett. {\bf 42}, 65 (1979).

\bibitem{sheehy_feshbach}
D.~E. Sheehy and L.~Radzihovsky,
\newblock Phys. Rev. Lett. {\bf 95}, 130402 (2005).

\end{thebibliography}

\end{document}